\documentclass[10pt,journal,twoside,web]{ieeecolor}
\usepackage{tmi2}
\usepackage{cite}
\usepackage{amsmath,amssymb,amsfonts}
\usepackage{algorithmic}
\usepackage{graphicx}
\usepackage{textcomp}
\def\BibTeX{{\rm B\kern-.05em{\sc i\kern-.025em b}\kern-.08em
    T\kern-.1667em\lower.7ex\hbox{E}\kern-.125emX}}
\usepackage{comment}
\usepackage{subfig}
\usepackage{tikz}
\usetikzlibrary{matrix,calc}
\usepackage{amsmath}
\usepackage{framed,multirow}
\usepackage{amssymb}
\usepackage{latexsym}
\usepackage{url}
\usepackage{xcolor}
\usepackage{lipsum}
\usepackage{hyperref}
\hypersetup{
    colorlinks=true,
    linkcolor=blue,
    filecolor=magenta,      
    urlcolor=cyan,
    pdftitle={MambaRoll},
    }
\usepackage{graphics}
\usepackage{gensymb}
\usepackage{caption}

\usepackage{soul}
\usepackage{amsfonts}
\usepackage{algorithmic}
\usepackage{graphicx}
\usepackage{textcomp}
\usepackage{booktabs}

\usepackage{bbding}

\usepackage{xcolor}
\definecolor{customblue}{RGB}{0, 153, 204}

\definecolor{newcolor}{rgb}{.8,.349,.1}
\def\baselinestretch{1.0}
\renewcommand{\arraystretch}{1.40}

\makeatletter
\IEEEtriggercmd{\reset@font\normalfont\fontsize{7.9pt}{8.40pt}\selectfont}
\makeatother
\IEEEtriggeratref{1}

\begin{document}
\title{Physics-Driven Autoregressive State Space Models for Medical Image Reconstruction}
\author{Bilal Kabas, Fuat Arslan, Valiyeh A. Nezhad, Saban Ozturk, Emine U. Saritas, and Tolga \c{C}ukur$^*$ \vspace{-1.2cm}
\\
\thanks{This study was supported in part by TUBA GEBIP 2015 and BAGEP 2017 fellowships, and by a TUBITAK 1001 Grant 121E488 awarded to T. \c{C}ukur (Corresponding author: Tolga \c{C}ukur, cukur@ee.bilkent.edu.tr).}
\thanks{B. Kabas, F. Arslan, V.A. Nezhad, S. Ozturk, E.U. Saritas, and T. \c{C}ukur are with the Dept. of Electrical and Electronics Engineering, and National Magnetic Resonance Research Center (UMRAM), Bilkent University, Ankara, Turkey (e-mails: \{bilal.kabas, fuat.arslan, valiyeh.ansarian, saban.ozturk\}@bilkent.edu.tr, saritas@ee.bilkent.edu.tr).}
}

\maketitle
\begin{abstract}
Medical image reconstruction from undersampled acquisitions is an ill-posed inverse problem requiring accurate recovery of anatomical structures from incomplete measurements. Physics-driven (PD) network models have gained prominence for this task by integrating data-consistency mechanisms with learned priors, enabling improved performance over purely data-driven approaches. However, reconstruction quality still hinges on the network’s ability to disentangle artifacts from true anatomical signals—both of which exhibit complex, multi-scale contextual structure. Convolutional neural networks (CNNs) capture local correlations but often struggle with non-local dependencies. While transformers aim to alleviate this limitation, practical implementations involve design compromises to reduce computational cost by balancing local and non-local sensitivity, occasionally resulting in performance comparable to CNNs. To address these challenges, we propose MambaRoll, a novel physics-driven autoregressive state space model (SSM) for high-fidelity and efficient image reconstruction. MambaRoll employs an unrolled architecture where each cascade autoregressively predicts finer-scale feature maps conditioned on coarser-scale representations, enabling consistent multi-scale context propagation. Each stage is built on a hierarchy of scale-specific PD-SSM modules that capture spatial dependencies while enforcing data consistency through residual correction. To further improve scale-aware learning, we introduce a Deep Multi-Scale Decoding (DMSD) loss, which provides supervision at intermediate spatial scales in alignment with the autoregressive design. Demonstrations on accelerated MRI and sparse-view CT reconstructions show that MambaRoll consistently outperforms state-of-the-art CNN-, transformer-, and SSM-based methods.
\vspace{-0.2cm}
\end{abstract}

\begin{IEEEkeywords}
medical image reconstruction, state space, autoregressive, physics-driven \vspace{-2mm}
\end{IEEEkeywords}

\bstctlcite{IEEEexample:BSTcontrol}

\begin{figure*}[t]
    \begin{minipage}{0.3\textwidth}
        \caption{Performance versus efficiency in MRI (left panel) and CT (right panel) reconstructions. Each method is represented by a circular marker whose center denotes the performance-latency trade-off (where latency is taken as inference time), and whose diameter indicates the memory load (see scale bars).}
        \label{fig:compute}
    \end{minipage}
    \begin{minipage}{0.7\textwidth}
        \centerline{\includegraphics[width=0.95\columnwidth]{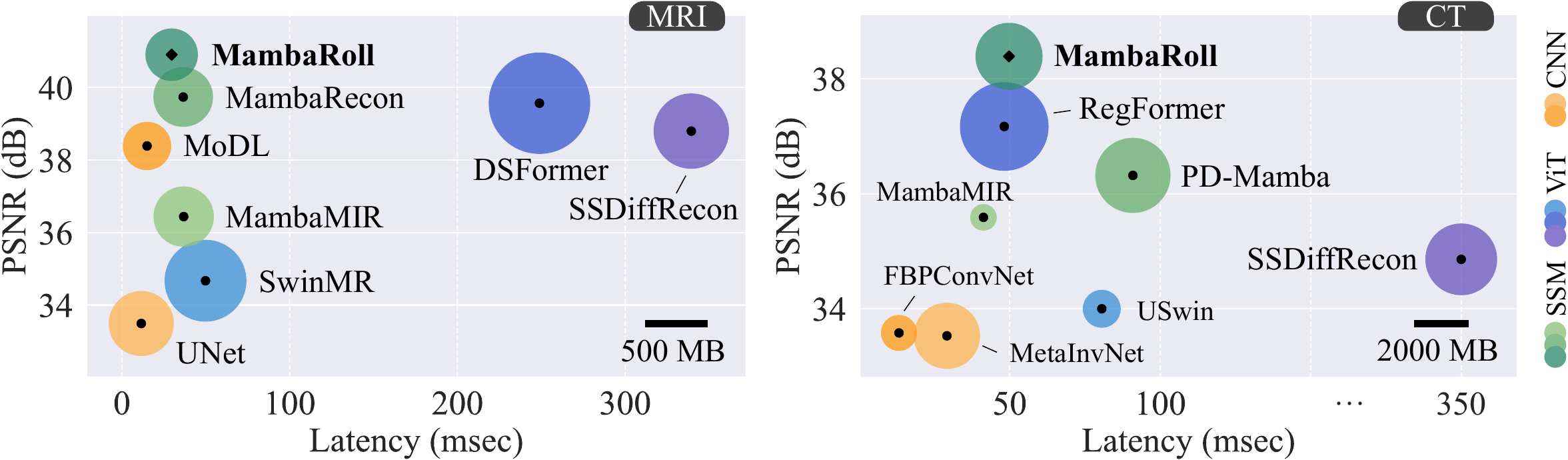}}
    \end{minipage}
    \hfill
\end{figure*}

\section{Introduction}
In many tomographic modalities, tissue structure is spatially encoded prior to acquisition, placing raw data in a distinct measurement domain (e.g., k-space in MRI, projection domain in CT). Higher spatial resolution requires collecting a greater number of encodings, but this increases scan time, patient discomfort, and/or radiation dose. A common strategy to improve scan efficiency is undersampling—acquiring only a subset of k-space points or projection views \cite{MIR-survey1}. However, reconstructing images from undersampled data is inherently ill-posed, requiring inversion of the imaging operator linking measurement and image domains \cite{sotos2}. Traditional methods often yield residual aliasing artifacts and noise during inversion \cite{MIR-survey3}, highlighting the need for robust reconstruction techniques.

Among learning-based reconstruction methods, data-driven (DD) approaches train networks to directly map input images linearly recovered from undersampled data to ground-truth images from fully-sampled acquisitions \cite{Wang2016,jongchul2018magnitude,han2018unet_ct,drone,lee2020sparse}. While DD methods build conditional priors that benefit from task-specific feature extraction, ignoring the imaging operator can impair performance \cite{cao2018CTuncond}.
In comparison, physics-driven (PD) approaches employ joint objectives based on learned network modules for artifact suppression and data-consistency steps guided by the imaging operator, and they often demonstrate stronger performance \cite{Schlemper2017,Hammernik2017,DreamNet,bindong2021,Zhou_2020_CVPR,akcakaya2020dense}. Unconditional PD variants (e.g., generative models) typically require intensive sampling at inference and are relatively difficult to deploy in resource-constrained clinical settings \cite{jalaln2021nips,gungor2022adaptive}. Instead, conditional PD approaches that train unrolled architectures kept frozen at test time offer a practical alternative \cite{unser2017,Yu2018c,Mardani2019b,MoDl}. 

While the PD framework is analytically well established in image reconstruction, its empirical success hinges on the specific design of network modules used to suppress aliasing artifacts. Critically, accurate separation of artifacts from true anatomical features demands sensitivity to contextual relationships in medical images across multiple spatial scales \cite{guo2022reconformer,korkmaz2022unsupervised}. Convolutional neural networks (CNNs) are adept at capturing local patterns but often struggle with long-range dependencies \cite{Schlemper2017,dar2020transfer,yang2021Fistanet}. Transformers offer improved modeling of non-local context, yet their substantial computational cost frequently necessitates simplified designs that compromise either spatial resolution or contextual expressiveness \cite{guo2022reconformer,wang2021dudotrans,RegFormer,huang2022swin,zhou2022dsformer,feng2023mtrans}. These trade-offs underscore the continued need for architectures that more effectively balance contextual sensitivity and computational efficiency \cite{wang2022bjork,ssdiffrecon}.

State space models (SSMs) emerge as a promising alternative in this regard \cite{zhu2024vision}. SSMs use rectilinear scans to rasterize images into one-dimensional (1D) sequences and model inter-pixel dependencies as a linear dynamical system with learnable parameters. This formulation promises stronger long-range sensitivity than CNNs and higher efficiency than transformers. Initial works have introduced SSMs in DD pipelines for medical image reconstruction \cite{mambamir,mmrmamba}, where neglect of the imaging physics can limit generalization across different operators. Only very recently, a PD architecture has been explored for MRI reconstruction \cite{mambarecon}. However, existing SSM methods rely on single-spatial-scale processing that can exhibit suboptimal sensitivity to distant pixel interactions, as such pixels may be separated by many intermediate elements when converted to 1D sequences \cite{liu2024vmamba}, compromising comprehensive contextual feature capture across spatial scales.

Given that many existing reconstruction models use network backbones that process multi-scale feature representations (e.g., UNet, Swin Transformer), extending SSMs to operate on multi-scale feature maps could alleviate limitations elicited by single-scale processing. However, without explicit cross-scale linking mechanisms and supervision of intermediate representations, multi-scale feature maps alone may be insufficient for high-fidelity reconstruction. Naive linking strategies can lead to inconsistencies and artifacts degrading image quality \cite{mustGAN}, while conventional supervision focused solely on final reconstructed images might provide suboptimal guidance for learning accurate intermediate-scale representations \cite{Goodfellow2016}.

\begin{figure*}[t]
    \begin{minipage}{0.225\textwidth}
       \captionsetup{justification=justified, singlelinecheck=false}
        \caption{MambaRoll employs an unrolled architecture with $K$ cascades to map a linear reconstruction $\hat{x}_0$ to the learned output $\hat{x}_K$. Each cascade progressively recovers high-resolution contextualized features $g_S$ across $S$ spatial scales via autoregressive prediction (Fig. \ref{fig:autoregressive}). At scale $s$, the proposed PD-SSM module convolutionally encodes input feature maps $f_{<s}$, captures contextualized feature representations through a compressed SSM block, decodes high-resolution feature maps, and residually enforces data fidelity. Following the final PD-SSM module within a cascade, a refinement module further restores fine textures.}
        \label{fig:mambaroll}
    \end{minipage}
    \begin{minipage}{0.775\textwidth}
        \centerline{\includegraphics[width=0.99\columnwidth]{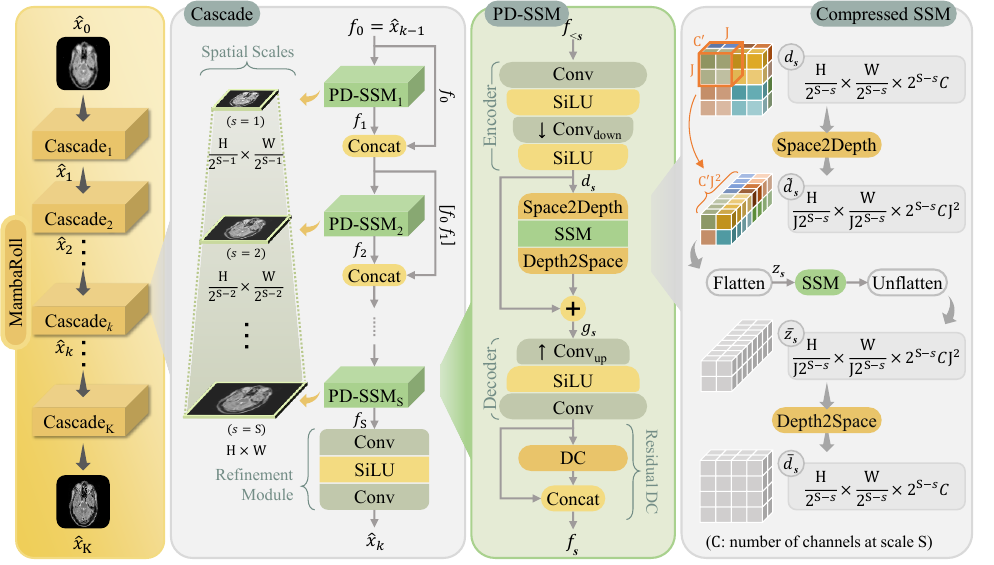}}
    \end{minipage}
    \hfill
   \end{figure*}

To address these challenges, we introduce MambaRoll, a novel physics-driven autoregressive SSM for high-fidelity and efficient medical image reconstruction (Fig. \ref{fig:compute}). For this purpose, MambaRoll uniquely integrates three key advances: (i) Multi-scale state space modeling, where SSMs operate across hierarchical feature resolutions to overcome the limitations of single-scale modeling; (ii) Autoregressive linking across scales, where finer features conditionally refine coarse-scale predictions, encouraging consistent anatomical structure; and (iii) Physics-driven unrolled optimization, where each stage combines a learned PD-SSM module with operator-based data consistency steps and maintains high efficiency through compressed SSM blocks using space-to-depth transformation (Fig. \ref{fig:mambaroll}). \textit{To our knowledge, MambaRoll is the first framework to unify multi-scale SSMs, autoregressive modeling, and PD unrolled optimization in a single architecture.}

To further enhance multi-scale learning, we introduce a Deep Multi-Scale Decoding (DMSD) loss that supervises intermediate reconstructions at each scale. This hierarchical loss structure provides direct feedback to each prediction stage, in contrast to conventional training strategies that only supervise final outputs.  Comprehensive experiments on accelerated MRI and sparse-view CT demonstrate that MambaRoll surpasses state-of-the-art DD and PD methods based on CNN, transformer, or SSM backbones. Code for MambaRoll is publicly available at {\small \url{https://github.com/icon-lab/MambaRoll}}.

\vspace{0.1cm}
\subsubsection*{\textbf{Contributions}}
\begin{itemize}
\item We introduce MambaRoll, a novel physics-driven SSM that enhances contextual sensitivity and reconstruction fidelity through multi-scale autoregressive inference.

\item We design an unrolled architecture that predicts feature maps at progressively finer spatial scales, with each level conditioned on coarser-scale outputs to support hierarchical context integration.

\item The architecture comprises scale-specific PD-SSM modules that maintain efficiency via compressed SSM blocks, and data consistency via corrections at the image scale.

\item We propose a DMSD loss that supervises intermediate representations at each spatial scale to improve overall reconstruction accuracy.

\end{itemize}
 
\section{Related Work}

\textbf{State Space Models:}
SSMs have emerged as an efficient alternative to transformers with refined balance between local and non-local contextual sensitivity \cite{mambamir}. \textit{Building on this strength, we introduce MambaRoll as the first model combining multi-scale SSMs, autoregressive modeling and PD unrolled optimization for medical image reconstruction.} Several earlier studies have explored SSMs for MRI reconstruction \cite{mambamir, mmrmamba, mambarecon}, with only MambaMIR extending to CT and PET \cite{mambamir}. These works demonstrate SSM's potential in image reconstruction but typically employ either DD (MambaMIR and MMR-Mamba) \cite{mambamir, mmrmamba} or PD (MambaRecon) \cite{mambarecon} designs based on single-spatial-scale SSM blocks, potentially causing suboptimal capture of multi-scale contextual features in medical images. In contrast, MambaRoll introduces PD-SSM modules operating progressively over spatial scales, enabling explicit linking of multi-scale contextual features.

After our preprint release \cite{mambaroll}, we became aware of two concurrent efforts exploring SSMs for MRI reconstruction. While these highlight growing interest in SSMs, MambaRoll carries principal novelties in architecture and application scope. DH-Mamba proposes a DD SSM method with hierarchically tokenized scan trajectories to enhance contextual modeling \cite{dmmamba}. Unlike DH-Mamba, MambaRoll employs a PD approach and explicitly integrates autoregressive priors across spatial scales to link multi-scale feature maps. CIMP-Net devises a PD method with spatial and spectral SSM modules for improved global context modeling \cite{causalmamba}. Yet, both SSM modules are confined to single spatial scales, potentially limiting contextual reach. MambaRoll instead introduces multi-scale PD-SSM modules and DMSD training loss to enhance sensitivity to multi-scale contextual features. Finally, while most prior SSM-based methods are limited to MRI, except MambaMIR, MambaRoll is designed to generalize to other medical image reconstruction tasks under a unified framework.

\textbf{Autoregressive Models:}
Autoregressive modeling decomposes joint probability distributions of element sequences into products of conditional distributions, such that each element is predicted conditioned on the preceding subset. Several previous CNN-based reconstruction studies have applied autoregressive modeling to capture dependencies between elements in raw pixel sequences \cite{luo2020pixelcnn,LORAKI} or between temporal steps in diffusion generative models \cite{luo2024autoregressive}. In contrast, MambaRoll introduces a PD framework integrating autoregressive modeling to capture dependencies between feature maps at different spatial scales, leveraging a novel SSM backbone. \textit{This represents, to the best of our knowledge, the first integration of spatial-scale autoregression in a medical image reconstruction model}. With these technical advances, we provide demonstrations of MambaRoll for performant MRI and CT reconstructions.

\section{Theory}
\subsection{Medical Image Reconstruction}
In many modalities, acquired data and the underlying image reflecting the spatial distribution of tissue structure are associated through a linear system \cite{MIR-survey1}:
\begin{equation}
        \label{eq:imop}
        \mathcal{A} x + \varepsilon = y ,
\end{equation}
where $x \in \mathbb{C}^{N}$ denotes the underlying complex image in vector form ($N$: number of pixels), $y \in \mathbb{C}^M$ are acquired complex data in vector form ($M$: number of measurements), $\varepsilon$ is measurement noise, and $\mathcal{A} \in \mathbb{C}^{M \times N}$ is the imaging operator that describes the relationship between image and measurement domains. For MRI scans, $\mathcal{A}=\mathcal{P}\mathcal{F}\mathcal{C}$ where $\mathcal{P}$ denotes the k-space sampling pattern, $\mathcal{F}$ denotes Fourier transformation, and $\mathcal{C}$ denotes coil sensitivities. Accordingly, $\mathcal{A}$ can be computed based on $\mathcal{P}$ and $\mathcal{F}$ that depend on known scan parameters, and $\mathcal{C}$ can be estimated from data in a central calibration region \cite{Hammernik2017}. For CT scans, $\mathcal{A}$ can be taken as a Radon transformation $\mathcal{R}$ \cite{unser2017}. Image reconstruction from acquired measurements involves solution of Eq. \ref{eq:imop} by inverting $\mathcal{A}$, which becomes increasingly ill-conditioned as the number of measurements decline with respect to the number of image pixels \cite{MIR-survey1}. Thus, image priors are employed to improve problem conditioning by regularizing reconstructions:
\begin{equation}
        \label{eq:reconstruction}
        \hat{x} = \arg\min_{x} \,\, \left\| \mathcal{A} x - y \right\|^2_2 + R(x),
\end{equation}
where $R(x)$ is an image prior that regularizes reconstructions based on the distribution of high-quality medical images. 

In PD models, the solution of Eq. \ref{eq:reconstruction} is typically operationalized as iterations through cascades of an unrolled architecture $G_{\theta_k}$, such that the output of cascade $k \in [1\,\,K]$ depends on the output of the previous cascade \cite{MoDl}:
\begin{equation}
        \label{eq:unrolled}
        \hat{x}_{k} = G_{\theta_k}(\hat{x}_{k-1};\mathcal{A},y),
\end{equation}
where $\theta_k$ denotes the parameters of the $k$th cascade, and $\hat{x}_{0} = \mathcal{A}^{\dagger}y$ is a linear reconstruction of undersampled data with $\dagger$ denoting the Hermitian adjoint. For a given cascade, $G_{\theta_k}(\hat{x}_{k-1};\mathcal{A},y) := \Psi_{dc}(\Psi_{nm}(\hat{x}_{k-1});\mathcal{A},y)$ interleaves projections through network and data-consistency (DC) modules:
\begin{align}
    v_{out}&=\Psi_{nm}(v_{in}) = \arg\min_{v_{out}} R_{\theta_k}(v_{out}|v_{in}), \label{eq:projR} \\
    v_{out}&=\Psi_{dc}(v_{in};A,y) = v_{in} + \mathcal{A}^{\dagger}(y - \mathcal{A} v_{in}), \label{eq:projDC} 
\end{align}
where the projection in Eq. \ref{eq:projR} is a forward pass through the network module, $v_{in}$ and $v_{out}$ denote the projection input and output, respectively. Performance in PD models depends critically on the ability of $\Psi_{nm}$ in separating undersampling-related artifacts from underlying tissue structure.

\subsection{MambaRoll}
MambaRoll employs PD-SSM modules to recover high-resolution medical images gradually across spatial scales, and a refinement module (RM) to enhance recovery of fine-grained tissue structure. Inspired by recent advances in autoregressive modeling for computer vision \cite{tian2024VAR}, we design PD-SSMs to autoregressively predict feature maps at finer spatial scales conditioned on coarser-scale representations, enhancing contextual representation by progressively aggregating multi-scale features while remaining consistent with acquired data.

\subsubsection{Network Architecture}
\begin{figure}[t]
       \centering
       \includegraphics[width=0.845\linewidth]{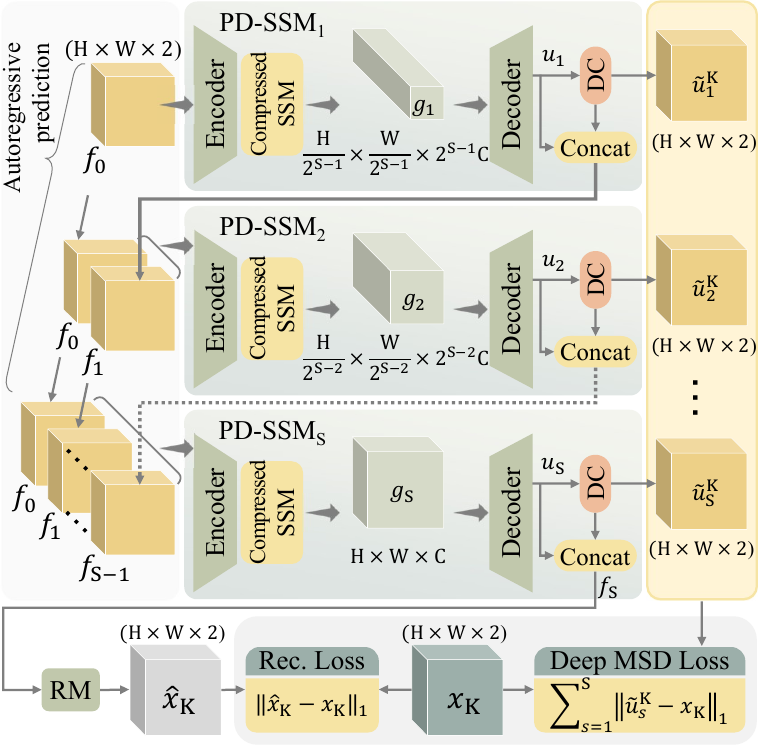}
       \caption{To explicitly link multi-scale representations via an autoregressive prior, the PD-SSM module at a given scale $s$ predicts the finer-scale feature map ($f_s$) conditioned on the preceding set of feature maps obtained at coarser scales ($f_{< s}:=[f_{0} \,\, f_{1} \,\, ... \,\, f_{s-1}]$). For model supervision, a deep multi-scale decoding loss (DMSD) loss is used in combination with a conventional reconstruction loss, where $\hat{x}_K$ is derived via the refinement module (RM).} 
       \label{fig:autoregressive}
   \end{figure}

Receiving a linear reconstruction of undersampled data as a two-channel input $\hat{x}_{0} = \mathcal{A}^{\dagger}y \in \mathbb{R}^{H \times W \times 2}$ with channels storing real and imaginary components ($H$: image height, $W$: image width), MambaRoll projects its input through $K$ cascades comprising \underline{\textit{novel PD-SSM modules}} and refinement modules to compute the reconstruction $\hat{x}_{K} \in \mathbb{R}^{H \times W \times 2}$. At cascade $k$, the input image $f_0 = \hat{x}_{k-1} \in \mathbb{R}^{H \times W \times 2}$ is projected through $S$ consecutive PD-SSM modules that autoregressively process feature maps across multiple spatial scales (Fig. \ref{fig:mambaroll}):  
\begin{equation}
f_s = \mathrm{PD}\mathrm{-}\mathrm{SSM}_s(f_{< s}), \,\,\,\, \mathrm{for} \,\,\, s \in \{1, 2, ..., S\},
\end{equation}
where $s$ denotes the scale index, $f_s \in \mathbb{R}^{H \times W \times 4}$ for $s \geq 1$ (4 channels due to residual DC blocks expressed in Eq. \ref{eq:resDC}), and $f_{< s}:=[f_{0} \,\, f_{1} \,\, ... \,\, f_{s-1}] \in \mathbb{R}^{H \times W \times (4s-2)}$ is formed by concatenating feature maps from PD-SSMs at earlier scales. To ensure PD-SSMs process feature maps while maintaining data fidelity, $f_s $ for all $s$ carry the same spatial dimensionality as the input image $\hat{x}_{0}$. \textit{Thus, processing at different spatial scales is achieved by adjusting resolution internally within PD-SSMs (Fig. \ref{fig:autoregressive})}. Afterwards, the RM module enhances recovery of structural details via convolutional layers: 
\begin{equation}
\hat{x}_k = \mathrm{RM}(f_S) := \mathrm{Conv}(\sigma(\mathrm{Conv}(f_S))),
\end{equation}
where $\hat{x}_k \in \mathbb{R}^{H \times W \times 2}$ is the output of the $k$th cascade, and $\sigma$ is a SiLU activation function.

\textbf{PD-SSM modules.} Diverging from conventional SSM modules, PD-SSMs enable progressive image recovery across multiple spatial scales while maintaining data fidelity at each scale. A given PD-SSM$_s$ module at scale $s$ first projects its input $f_{< s} \in \mathbb{R}^{H \times W \times 4s-2}$ through a convolutional encoder to map the high-resolution input onto the desired scale:
\begin{equation}
    d_s = \mathrm{Enc}_s (f_{< s}) := \sigma(\mathrm{Conv}_\mathrm{down}(\sigma(\mathrm{Conv}(f_{< s})))),
\end{equation}
where $d_s \in \mathbb{R}^{H_s \times W_s \times C_s}$ with $H_s=H/2^{(S-s)}$, $W_s=W/2^{(S-s)}$, $C_s=2^{(S-s)}C$ ($C$: number of channels at scale $S$), and $\mathrm{Conv}_\mathrm{down}$ performs spatial downsampling while $\mathrm{Conv}$ expands channel dimensionality to $C_s$. 

A compressed SSM block with residual connection extracts contextual features at the $s$th scale as $g_s = \mathrm{SSM}(d_s)$, while improving efficiency in state-space modeling. Since conventional SSMs' computational complexity scales with sequence length (number of image pixels), our compressed SSM blocks apply space-to-depth transformation on input feature map $d_s$ to shorten the sequence length by factor $J^2$, tiling $J \times J$ pixel patches along the channel dimension \cite{pixelshuffle}:
\begin{align}
&\Tilde{d}_s[w_1, w_2,\Tilde{c}] = d_s[J(w_1-1) + v_1, J(w_2-1) + v_2,c],\\
&c = \left\lceil \frac{\Tilde{c}}{J^2} \right\rceil, \quad
v_1 = \left\lceil \frac{((\Tilde{c}-1) \bmod J^2) + 1}{J}) \right\rceil,\nonumber\\
&v_2 = ((\Tilde{c}-1) \bmod J)+1.\nonumber
\end{align}
where $w_1 \in \{1,...,H_s/J\}$, $w_2 \in \{1,...,W_s/J\}$, and $\Tilde{c} \in \{1,...,C_sJ^2\}$. The resultant feature map $\Tilde{d}_s \in \mathbb{R}^{H_s/J \times W_s/J \times C_{s}J^2}$ is transformed onto a sequence $z_{s} \in \mathbb{R}^{H_s W_s/J^2 \times C_sJ^2}$ via a two-dimensional sweep scan, and independently processed across channels with a discretized SSM:  
\begin{align}
    h_s[n,c] &= \mathbf{A} h_s[n-1,c] + \mathbf{B} z_s[n,c], \label{eq:ssm1} \\
    \bar{z}_s[n,c] &= \mathbf{C} h_s[n,c], \label{eq:ssm2}
\end{align}
where $n \in [1 \mbox{ } H_s W_s/J^2]$ denotes sequence index, \(h_s[n,c] \in \mathbb{R}^{D \times 1}\) denotes the scale-specific hidden state at step $n$, and \(\mathbf{A} \in \mathbb{R}^{D \times D}\), \(\mathbf{B} \in \mathbb{R}^{D \times 1}\), \(\mathbf{C} \in \mathbb{R}^{1 \times D}\) are learnable parameters (\(D\): the state dimensionality). After the SSM layer, the output sequence  $\bar{z}_s$ is expanded and depth-to-space transformed:
\begin{align}
&\bar{d}_s[J(w_1-1) + v_1, J(w_2-1) + v_2,c] = \mathrm{expand}(\bar{z}_s)[w_1, w_2, \Tilde{c}],\\
&\Tilde{c} = (c-1) \cdot J^2 + (v_1-1) \cdot J + v_2, \quad v_1, v_2 \in \{1, ..., J\}.\nonumber
\end{align}
The restored feature map $\bar{d}_s \in \mathbb{R}^{H_s \times W_s \times C_s}$ is residually added onto the input:
\begin{equation}
    g_s = \bar{d}_s + d_s.
\end{equation}

To enable PD-SSM modules to enforce fidelity to acquired data, the contextualized feature map $g_s$ is projected through a convolutional decoder to recollect a high-resolution image:
\begin{equation}
   u_s = \mathrm{Dec}_s(g_s) := \mathrm{Conv}(\sigma(\mathrm{Conv}_\mathrm{up}(g_s))),
\end{equation}
where $u_s \in \mathbb{R}^{H \times W \times 2}$, $\mathrm{Conv}_\mathrm{up}$ performs spatial upsampling while $\mathrm{Conv}$ reduces channel dimensionality to 2 to store real and imaginary components. Finally, the recollected image can be projected through a residual data-consistency (DC) block that concatenates $u_s$ with a data-consistent version of itself:
\begin{align}
     \label{eq:resDC}
     \tilde{u}_s &=  \mathrm{DC}(u_s) := u_s + \mathcal{A}^{\dagger}(y - \mathcal{A} u_s)), \\
     f_s &= \mathrm{resDC}(u_s) := [u_s \,\,\,\, \tilde{u}_s],
\end{align}
where $\tilde{u}_s$ ensures that reconstructed data are consistent with acquired data $y$, and $u_s$ allows fine-grained control over the emphasis on data consistency to optimize task performance.

 \begin{figure*}[t]
    \begin{minipage}{0.79\textwidth}
        \centerline{\includegraphics[width=0.97\columnwidth]{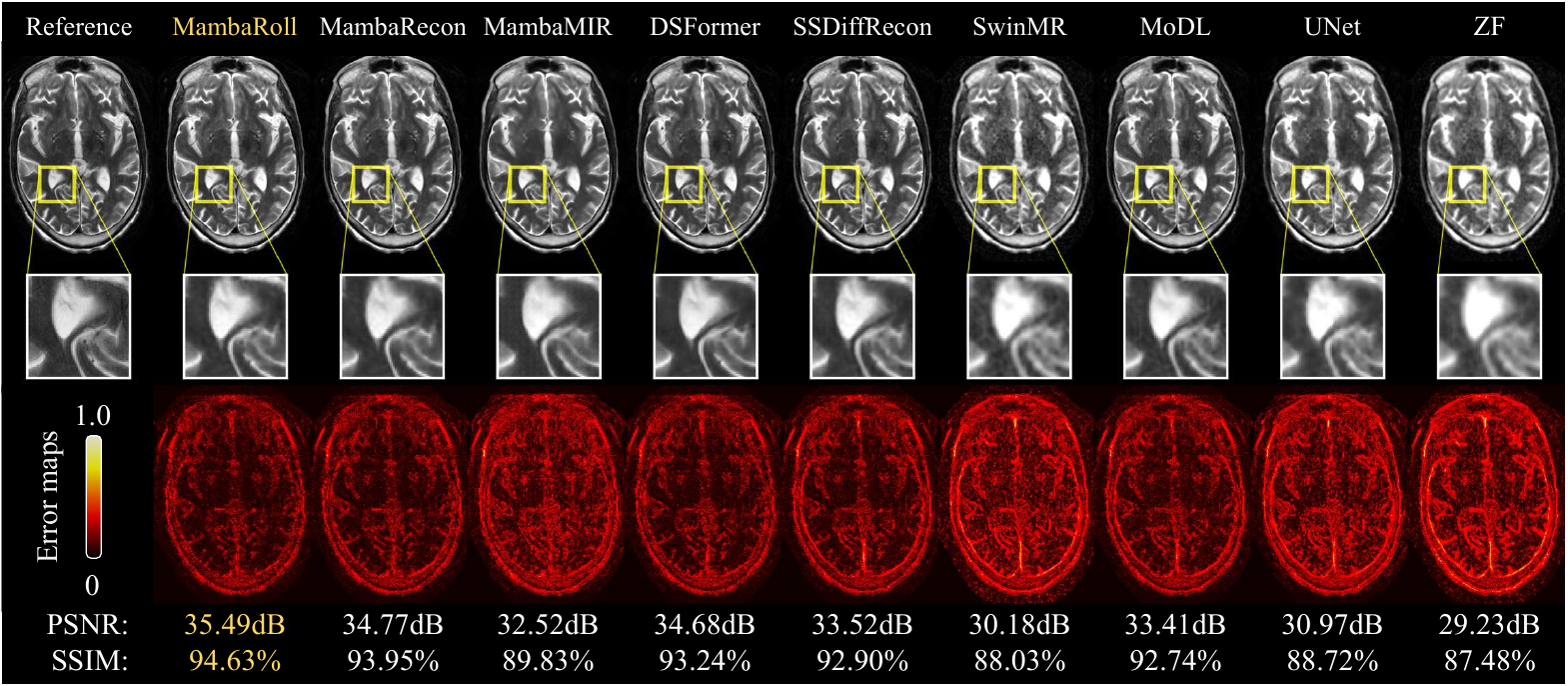}}
    \end{minipage}
    \begin{minipage}{0.20\textwidth}
        \caption{Accelerated MRI reconstructions of a representative cross-section from a T\textsubscript{2}-weighted acquisition at $R=12$. Reconstructed images from competing methods are shown along with a linear reconstruction of undersampled data (ZF), and the reference image derived from fully-sampled data. Zoom-in windows and error maps are included to highlight differences in tissue depiction. PSNR and SSIM values are listed below each reconstruction.}
        \label{fig:fastmri}
    \end{minipage}
    \hfill
\end{figure*}

\begin{table*}[t]
    \centering
    \footnotesize
    \caption{MRI reconstruction performance at $R= 4-12$. PSNR (dB) and SSIM (\%) listed as mean$\pm$std across the test set. Boldface marks the top performing method. (DD: data-driven, PD: physics-driven, CNN: convolutional, ViT: transformer, SSM: state-space)}
     \renewcommand{\arraystretch}{1.1}
      \resizebox{0.9\textwidth}{!}{
        \begin{tabular}{l|cc|ccc|cccccccc}
            \hline
            \multicolumn{1}{c}{} & \multicolumn{2}{|c|}{Framework} & \multicolumn{3}{c|}{Architecture} & \multicolumn{2}{c}{$R = 4$} & \multicolumn{2}{c}{$R = 8$} & \multicolumn{2}{c}{$R = 12$} & \multicolumn{2}{c}{Average}\\ 
            \cline{2-14}
            \multicolumn{1}{c}{} & \multicolumn{1}{|c}{DD} & \multicolumn{1}{c|}{PD} & \multicolumn{1}{c}{CNN} & \multicolumn{1}{c}{ViT} & \multicolumn{1}{c|}{SSM} & PSNR & SSIM & PSNR & SSIM & PSNR & SSIM & PSNR & SSIM\\ 
            \hline
            UNet \cite{patelFLMRI} & \Checkmark & & \Checkmark & & & 36.25$\pm$2.42 & 94.27$\pm$3.53 & 33.93$\pm$2.38 & 91.63$\pm$4.98 & 30.29$\pm$5.43 & 89.63$\pm$5.69 & 33.49 & 91.84 \\ \hline
            MoDL \cite{MoDl} & &\Checkmark & \Checkmark& & & 41.76$\pm$2.55 & 97.64$\pm$2.08 & 37.56$\pm$2.60 & 95.03$\pm$3.69 & 35.87$\pm$2.41 & 93.19$\pm$4.48 & 38.39 & 95.28 \\ \hline
            SwinMR \cite{huang2022swin} &\Checkmark & & & \Checkmark & & 37.31$\pm$2.38 & 94.63$\pm$2.61 & 34.10$\pm$2.18 & 91.58$\pm$3.90 & 32.60$\pm$2.06 & 89.65$\pm$4.47 & 34.67 & 91.95 \\ \hline
            SSDiffRecon \cite{ssdiffrecon} & & \Checkmark & & \Checkmark & & 42.47$\pm$2.99 & 97.63$\pm$2.14 & 37.99$\pm$2.66 & 94.79$\pm$3.71 & 35.95$\pm$2.45 & 93.12$\pm$4.54 & 38.80 & 95.18 \\ \hline
            DSFormer \cite{zhou2022dsformer} & & \Checkmark & & \Checkmark & & 43.13$\pm$3.12 & 97.91$\pm$2.12 & 38.69$\pm$2.84 & 95.04$\pm$3.74 & 36.90$\pm$2.71 & 93.23$\pm$4.62 & 39.57 & 95.39 \\ \hline
            MambaMIR \cite{mambamir} & \Checkmark & & & & \Checkmark & 38.58$\pm$2.60 & 95.12$\pm$3.23 & 36.05$\pm$2.56 & 92.66$\pm$4.88 & 34.70$\pm$2.47 & 91.38$\pm$5.61 & 36.44 & 93.05 \\ \hline
            MambaRecon \cite{mambarecon} & & \Checkmark & & & \Checkmark & 43.45$\pm$3.31 & 98.16$\pm$2.31 & 38.74$\pm$2.85 & 95.57$\pm$3.94 & 37.03$\pm$2.72 & 93.99$\pm$4.69 & 39.74 & 95.91 \\ \hline
            MambaRoll &  & \Checkmark & & & \Checkmark & \textbf{45.55$\pm$3.71} & \textbf{98.57$\pm$3.57} & \textbf{39.51$\pm$3.04} & \textbf{96.12$\pm$3.71} & \textbf{37.68$\pm$2.77} & \textbf{94.61$\pm$4.37} & \textbf{40.91} & \textbf{96.43} \\ \hline
        \end{tabular}
      }
    \label{tab:fastmri}
\end{table*}

\subsubsection{Deep Multi-Scale Decoding (DMSD) Loss}
MambaRoll leverages autoregressive processing across spatial scales to capture multi-scale contextual features. Yet, processing accuracy at each scale depends crucially on its alignment with underlying anatomical structure. Thus, we introduce a DMSD loss to strengthen intermediate scale learning. The feature map $g_s^K$ at scale $s$ within the final cascade (i.e., $K$th) is decoded to a full-resolution image $u_s^K = \text{Dec}_s(g_s^K)$ through a learned decoder (Fig. \ref{fig:autoregressive}). These decoded images are then projected through the DC block, and supervised against the ground-truth image $x_K$ to enable direct learning at multiple scales:
\begin{equation}
\mathcal{L}_{\text{DMSD}} = \sum_{s=1}^{S} \| \tilde{u}_s^K - x_K \|_1 = \sum_{s=1}^{S} \| \text{DC}(\text{Dec}_s(g_s^K)) - x_K \|_1.
\label{eq:fmap}
\end{equation}
By supervising a strictly data-consistency enforced version of decoded images (i.e., $\tilde{u}_s^K$) rather than the residually-corrected images (i.e., $f_s^K$), the DMSD loss promotes formation of scale-specific features more closely aligned with the ground-truth image. The overall loss then combines the reconstruction loss on the predicted image with Eq. \ref{eq:fmap}:
\begin{equation}
\mathcal{L}_{\text{total}} = \| \hat{x}_K - x_K \|_1 + \mathcal{L}_{\text{DMSD}},
\end{equation}
where $\hat{x}_K$ denoted the model’s predicted image. DMSD loss departs from conventional supervision by treating scale-wise predictions not merely as intermediate latent steps, but as independently decodable image estimates. By grounding images decoded at separate spatial scales individually to ground-truth images, this approach helps mitigate residual artifacts within feature maps at coarser spatial scales.

\section{Methods}

\subsection{Datasets}

We conducted evaluations on two public datasets with random subject-level splits into independent training, validation and test sets. \textbf{Accelerated MRI} experiments used brain data from fastMRI \cite{fastmri}. T\textsubscript{1}, T\textsubscript{2} and FLAIR contrasts were analyzed as multi-coil complex k-space data, each volume comprising 10 cross-sections of size 320$\times$320. Multi-coil data were compressed onto 5 virtual coils preserving 90\% of variance \cite{zhang2013coil}, with coil sensitivities estimated via ESPIRiT \cite{uecker2014espirit}. Data were split into (240, 60, 120) subjects for training, validation, test sets, corresponding to (7200, 1800, 3600) cross-sectional samples. Two-dimensional variable-density undersampling at rates $R=4-12$ was employed \cite{lustig2007sparse}. Linear reconstructions were obtained by zero-filling missing k-space samples followed by inverse Fourier transformation (ZF). \textbf{Sparse-view CT} experiments used lung data from LoDoPaB-CT \cite{leuschner2021lodopab}. Each volume comprised 90 cross-sections of size 352$\times$352. Data were split into (30, 6, 12) subjects for training, validation, test sets, corresponding to (2700, 540, 1080) cross-sectional samples. For sparse-view CT scans, ground-truth reconstructions were Radon transformed using two-dimensional parallel-beam geometry, with sinograms uniformly undersampled at rates $R=4-8$ across the angular dimension. Linear reconstructions used filtered back projection (FBP).
 
\subsection{Architectural Details}
MambaRoll used $S=3$ spatial scales (at 0.25, 0.5, 1 scale of the original image resolution). In PD-SSM modules, the encoder used expansion factors across the channel dimension of (4, 2, 1), whereas the decoder used expansion factors of (0.25, 0.5, 1) across scales. The SSM block used a compression factor of $J=4$ and a sweep-scan trajectory to project feature maps onto a sequence, and a state-dimensionality of $D=16$. The refinement module comprised a convolutional block. All convolution layers used 3$\times$3 kernels and SiLU activation.

   \begin{figure*}[t]
        \begin{minipage}{0.79\textwidth}
            \centerline{\includegraphics[width=0.9\columnwidth]{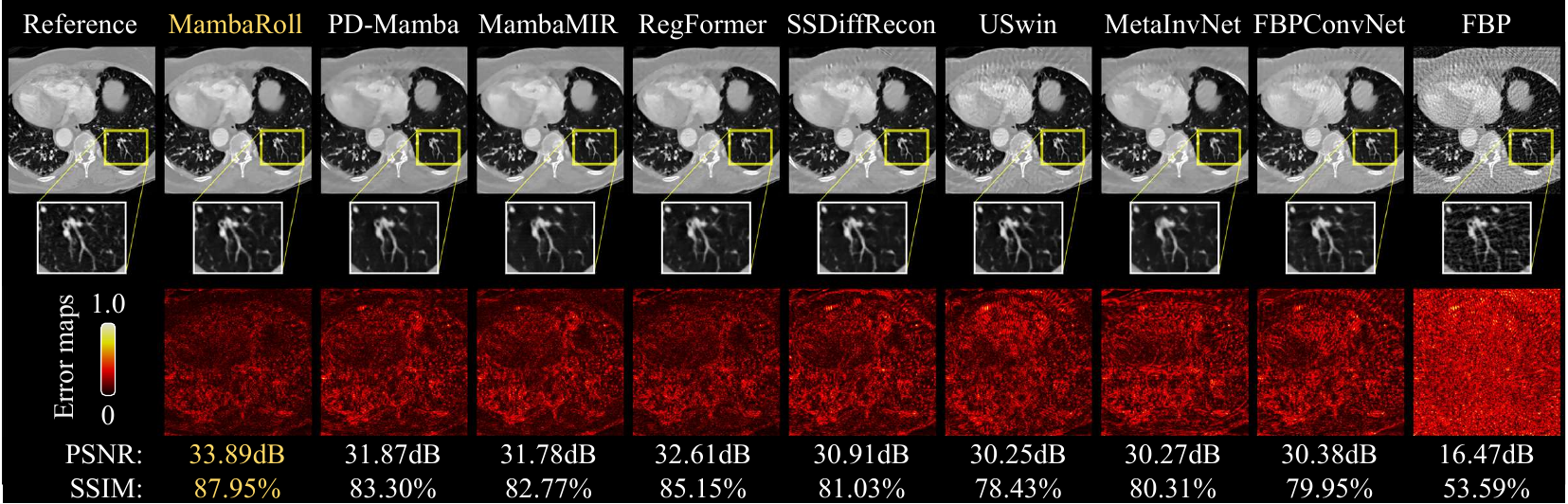}}
        \end{minipage}
        \begin{minipage}{0.20\textwidth}
            \caption{Sparse-view CT reconstructions of a representative cross-section at $R=8$. Reconstructed images from competing methods are shown along with a linear reconstruction of undersampled data (FBP), and the reference image derived from fully-sampled data.}
            \label{fig:ct}
        \end{minipage}
        \hfill
    \end{figure*}

       \begin{table*}[t]
        \centering
        \footnotesize
        \caption{CT reconstruction performance at $R= 4-8$. Boldface marks the top performing method.}
        \renewcommand{\arraystretch}{1.1}
         \resizebox{0.9\textwidth}{!}{
        \begin{tabular}{l|cc|ccc|cccccccc}
            \hline
            \multicolumn{1}{c}{} & \multicolumn{2}{|c|}{Framework} & \multicolumn{3}{c|}{Architecture} & \multicolumn{2}{c}{$R = 4$} & \multicolumn{2}{c}{$R = 6$} & \multicolumn{2}{c}{$R = 8$} & \multicolumn{2}{c}{Average}\\ 
            \cline{2-14}
            \multicolumn{1}{c}{} & \multicolumn{1}{|c}{DD} & \multicolumn{1}{c|}{PD} & \multicolumn{1}{c}{CNN} & \multicolumn{1}{c}{ViT} & \multicolumn{1}{c|}{SSM} & PSNR & SSIM & PSNR & SSIM & PSNR & SSIM & PSNR & SSIM\\ 
            \hline
            FBPConvNet \cite{unser2017} & \Checkmark &  & \Checkmark & &  & 34.41$\pm$1.65 & 85.68$\pm$4.92 & 33.85$\pm$1.73 & 84.86$\pm$5.75 & 32.49$\pm$1.57 & 81.76$\pm$5.94 & 33.58 & 84.10 \\ \hline
            MetaInv-Net \cite{bindong2021}  &  & \Checkmark & \Checkmark & &  & 36.31$\pm$2.23 & 90.03$\pm$5.72 & 33.10$\pm$1.68 & 84.82$\pm$6.61 & 31.17$\pm$1.38 & 80.90$\pm$6.82 & 33.53 & 85.25 \\ \hline
            USwin \cite{xu2024hybrid}   & \Checkmark &  & & \Checkmark &  & 36.41$\pm$2.58 & 91.96$\pm$5.33 & 33.48$\pm$1.94 & 87.41$\pm$6.24 & 32.11$\pm$1.68 & 84.21$\pm$6.48 & 34.00 & 87.86 \\ \hline
            SSDiffRecon \cite{ssdiffrecon} & & \Checkmark & & \Checkmark & & 38.25$\pm$2.75 & 91.95$\pm$4.96 & 34.92$\pm$2.11 & 86.97$\pm$6.01 & 31.40$\pm$1.54 & 80.99$\pm$7.19 & 34.86 & 86.64 \\ \hline
            RegFormer  \cite{RegFormer} &  & \Checkmark & & \Checkmark &  & 39.88$\pm$3.23 & 93.33$\pm$4.89 & 36.85$\pm$2.58 & 89.49$\pm$5.99 & 34.79$\pm$2.17 & 86.29$\pm$6.39 & 37.17 & 89.70 \\ \hline
            MambaMIR  \cite{mambamir} & \Checkmark &  & & & \Checkmark & 37.42$\pm$2.40 & 89.57$\pm$4.77 & 35.42$\pm$2.20 & 86.73$\pm$5.89 & 33.92$\pm$1.99 & 83.74$\pm$6.19 & 35.59 & 86.68 \\ \hline
            PD-Mamba \cite{liu2024vmamba} &  & \Checkmark & & & \Checkmark & 38.99$\pm$2.97 & 92.64$\pm$5.08 & 35.90$\pm$2.33 & 88.35$\pm$6.18 & 34.06$\pm$1.99 & 85.27$\pm$6.65 & 36.32 & 88.75 \\ \hline
            MambaRoll  &  & \Checkmark & & & \Checkmark & \textbf{40.72$\pm$3.56} & \textbf{94.05$\pm$4.86} & \textbf{38.00$\pm$2.93} & \textbf{91.15$\pm$5.97} & \textbf{36.44$\pm$2.62} & \textbf{89.06$\pm$6.49} & \textbf{38.39} & \textbf{91.42} \\ \hline
        \end{tabular}
         }
        \label{tab:ct}
    \end{table*}   
\subsection{Competing Methods}
    MambaRoll was demonstrated against state-of-the-art baselines in MRI and CT reconstruction, including both DD and PD methods. \textbf{For accelerated MRI reconstruction}, competing methods included CNNs (data-driven \ul{UNet} \cite{patelFLMRI} and physics-driven \ul{MoDL} \cite{MoDl}); transformers (data-driven \ul{SwinMR} \cite{huang2022swin}, physics-driven \ul{SSDiffRecon} \cite{korkmaz2023selfsupervised} and \ul{DSFormer} \cite{zhou2022dsformer}); and SSMs (data-driven \ul{MambaMIR} \cite{mambamir}, physics-driven \ul{MambaRecon} \cite{mambarecon}). \textbf{For sparse-view CT reconstruction}, competing methods included CNNs (data-driven \ul{FBPConvNet} \cite{unser2017}, physics-driven \ul{MetaInv-Net} \cite{bindong2021}); transformers (data-driven \ul{USwin} \cite{xu2024hybrid}, physics-driven \ul{SSDiffRecon} \cite{korkmaz2023selfsupervised} and  \ul{RegFormer} \cite{RegFormer}); and SSMs (data-driven \ul{MambaMIR} \cite{mambamir}, physics-driven \ul{PD-Mamba} based on \cite{liu2024vmamba}).

 \subsection{Modeling Procedures}
Network models used two separate input-output channels to represent real and imaginary components of MR images, and a single input-output channel for CT images. Key model hyperparameters, including the number of cascades in unrolled architectures ($K$) and learning rate, were selected to maximize performance on the validation set. Accordingly, $K$$=$$5$ in MRI reconstruction and $K$$=$$3$ in CT reconstruction yielded near-optimal performance consistently across methods, whereas variable learning rates ranging in $10^{-6}$ to $10^{-3}$ were selected for individual methods. Modeling was performed via the PyTorch framework on an Nvidia RTX 4000 (Ada) GPU. Models were trained and tested on two-dimensional cross sections. Training was conducted using the Adam optimizer for 50 epochs, $(\beta_1, \beta_2) = (0.9, 0.999)$. Performance was assessed by peak signal-to-noise ratio (PSNR) and structural similarity index (SSIM) between recovered and reference images. Reference images were derived via Fourier reconstruction of fully-sampled acquisitions in MRI, and via filtered back-projection of fully-sampled sinograms in CT. Significance of performance differences was assessed via Wilcoxon signed-rank tests.

\section{Results}

\subsection{Accelerated MRI Reconstruction}
\label{sec:comparison_mri}
Demonstrations were first performed for accelerated MRI reconstruction. MambaRoll was compared against state-of-the-art models based on convolutional (UNet, MoDL), transformer (SwinMR, SSDiffRecon, DSFormer), and SSM (MambaMIR, MambaRecon) backbones. Reconstruction performances at undersampling rates $R=4-12$ are listed in Table \ref{tab:fastmri}. MambaRoll significantly outperforms competing methods across reconstruction tasks (p$<$0.05). On average across tasks, MambaRoll offers (PSNR, SSIM) improvements of 4.97dB, 2.87\% over convolutional; 3.23dB, 2.26\% over transformer; and 2.82dB, 1.95\% over SSM baselines. Note that PD methods (MambaRoll, MambaRecon, SSDiffRecon, DSFormer, MoDL) generally yield improved performance against DD methods (MambaMIR, SwinMR, UNet). These results indicate that incorporating PD-SSM modules enhances reliability in MRI reconstruction and that MambaRoll's autoregressive state-space architecture outperforms convolutional, transformer, and conventional SSM backbones.

Representative reconstructions are shown in Fig. \ref{fig:fastmri}. DD baselines (MambaMIR, SwinMR, UNet) exhibit pronounced noise, residual artifacts, and visible blur. Meanwhile, PD baselines show less noise but display distinct issues: DSFormer presents ringing artifacts, MambaRecon shows pixelation near high-intensity tissue regions, and SSDiffRecon and MoDL exhibit edge blurring. MambaRoll achieves enhanced tissue delineation with reduced noise and artifact levels, which can be attributed to the autoregressive state-space approach that improves capture of multi-scale contextual features.

   \begin{figure*}[t]
        \begin{minipage}{0.79\textwidth}
            \centerline{\includegraphics[width=0.9\columnwidth]{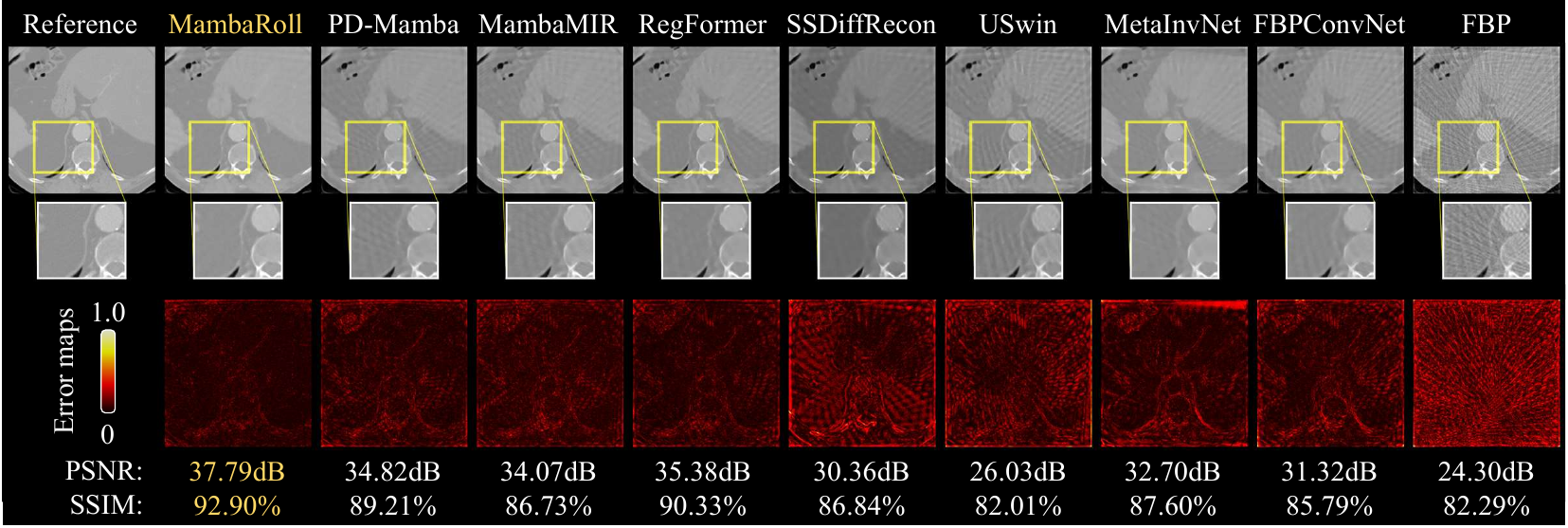}}
        \end{minipage}
        \begin{minipage}{0.20\textwidth}
            \caption{Sparse-view CT reconstructions of a representative cross-section at $R=8$ with models trained at $R=6$. Reconstructed images from competing methods are shown along with a linear reconstruction of undersampled data (FBP), and the reference image derived from fully-sampled data.}
            \label{fig:ctX}
        \end{minipage}
        \hfill
    \end{figure*} 
        
\subsection{Sparse-view CT Reconstruction}
   \label{sec:comparison_ct} 
Next, demonstrations were performed for sparse-view CT reconstruction. MambaRoll was compared against state-of-the-art models based on convolutional (FBPConvNet, MetaInv-Net), transformer (USwin, SSDiffRecon, RegFormer), and SSM (MambaMIR, PD-Mamba) backbones. Reconstruction performances at undersampling rates $R=4-8$ are listed in Table \ref{tab:ct}. MambaRoll significantly outperforms competing methods across reconstruction tasks (p$<$0.05). On average across tasks, MambaRoll offers (PSNR, SSIM) improvements of 4.84dB, 6.75\% over convolutional; 3.05dB, 3.35\% over transformer; and 2.44dB, 3.70\% over SSM baselines. While PD methods generally offer performance benefits over DD counterparts (e.g., PD-Mamba vs MambaMIR, RegFormer vs USwin), performance differences are moderate compared to MRI. MambaRoll's benefits over convolutional baselines are substantially stronger in CT, which might be attributed to the streaking-like aliasing artifacts spanning long distances in sparse-view CT images, compromising artifact suppression in convolutional methods with locality biases.

Representative images are displayed in Fig. \ref{fig:ct}. DD baselines (MambaMIR, USwin, FBPConvNet) show notable reconstruction errors due to spatial blurring and residual streaking artifacts. PD baselines (PD-Mamba, SSDiffRecon, RegFormer, MetaInvNet) yield improved performance but still suffer from moderate streaking artifacts that manifest as contrast inaccuracies near bright tissue signals. MambaRoll effectively mitigates streaking artifacts, yielding reconstructions that closely resemble ground truth in tissue depiction. These results indicate that MambaRoll's autoregressive state-space architecture improves capture of complex spatial dependencies, enabling enhanced artifact suppression and structural fidelity.

    \begin{table}[t]
        \centering
        \footnotesize
                \caption{Generalization performance in MRI reconstruction under variability in R values (R\textsubscript{training} $\rightarrow$ R\textsubscript{test}).} 
        \renewcommand{\arraystretch}{1.1}
         \resizebox{0.875\columnwidth}{!}{%
        \begin{tabular}{lccccc}
        \hline
        \multirow{2}{*}{} & \multicolumn{2}{c}{$R: 12 \rightarrow 8$} & \multicolumn{2}{c}{$R: 8 \rightarrow 12$} \\ \cline{2-5} & PSNR & SSIM & PSNR & SSIM \\ \hline
        UNet
            & 32.30$\pm$2.14 & 87.79$\pm$5.68
            & 31.31$\pm$1.98 & 86.84$\pm$6.07 \\ \hline
        MoDL
            & 36.41$\pm$2.54 & 92.95$\pm$4.00
            & 34.09$\pm$2.18 & 90.85$\pm$4.72 \\ \hline
        SwinMR
            & 33.70$\pm$2.02 & 90.45$\pm$4.33
            & 32.01$\pm$1.90 & 87.97$\pm$5.16 \\ \hline
        SSDiffRecon
            & 36.32$\pm$2.55 & 92.73$\pm$4.00
            & 34.44$\pm$2.29 & 90.69$\pm$4.84 \\ \hline
        DSFormer
            & 35.23$\pm$2.37 & 92.58$\pm$4.06
            & 35.00$\pm$2.44 & 91.02$\pm$4.76 \\ \hline
        MambaMIR
            & 32.95$\pm$2.21 & 88.98$\pm$5.34
            & 33.01$\pm$2.20 & 87.89$\pm$5.70 \\ \hline
        MambaRecon
            & 36.52$\pm$2.72 & 93.36$\pm$4.30
            & 35.45$\pm$2.49 & 91.99$\pm$4.60 \\ \hline
        MambaRoll 
            & \textbf{37.36$\pm$2.94} & \textbf{94.30$\pm$3.80}
            & \textbf{35.93$\pm$2.55} & \textbf{92.68$\pm$4.38} \\ \hline
        \end{tabular}
        }
        \label{tab:cross_mri}
    \end{table}
    
    \begin{table}[t]
        \centering
        \footnotesize
               \caption{Generalization performance in CT reconstruction under variability in R values (R\textsubscript{training} $\rightarrow$ R\textsubscript{test}).}
        \renewcommand{\arraystretch}{1.1}
       \resizebox{0.875\columnwidth}{!}{%
        \begin{tabular}{lccccc}
        \hline
        \multirow{2}{*}{} & \multicolumn{2}{c}{$R: 8 \rightarrow 6$} & \multicolumn{2}{c}{$R: 6 \rightarrow 8$} \\ \cline{2-5} & PSNR & SSIM & PSNR & SSIM \\ \hline
        FBPConvNet
            & 33.40$\pm$1.64 & 83.96$\pm$5.89
            & 31.29$\pm$1.48 & 79.44$\pm$6.21 \\ \hline
        MetaInvNet
            & 32.40$\pm$1.34 & 83.50$\pm$6.62
            & 30.72$\pm$1.46 & 79.96$\pm$6.68 \\ \hline
        USwin
            & 34.80$\pm$1.88 & 87.67$\pm$6.22
            & 28.91$\pm$1.94 & 78.04$\pm$6.55 \\ \hline
        SSDiffRecon
            & 31.71$\pm$1.40 & 81.92$\pm$7.16
            & 29.36$\pm$1.33 & 80.57$\pm$6.28 \\ \hline
        RegFormer
            & 35.75$\pm$2.18 & 88.39$\pm$6.01
            & 33.90$\pm$1.93 & 85.03$\pm$6.29 \\ \hline
        MambaMIR
            & 34.60$\pm$1.98 & 84.95$\pm$6.11
            & 32.80$\pm$1.67 & 82.29$\pm$5.87 \\ \hline
        PD-Mamba
            & 35.07$\pm$2.00 & 87.22$\pm$6.35
            & 33.38$\pm$1.84 & 84.05$\pm$6.42 \\ \hline
        MambaRoll
            & \textbf{37.53$\pm$2.76} & \textbf{90.75$\pm$6.11}
            & \textbf{34.68$\pm$2.37} & \textbf{85.81$\pm$6.34} \\ \hline
        \end{tabular}
         }
        \label{tab:cross_ct}
    \end{table}

\subsection{Generalization Under Acquisition Variability}
To investigate the robustness of reconstruction models under distributional shifts, we evaluated generalization performance across varying undersampling rates—an important aspect of domain variability in accelerated imaging. As representative scenarios, we considered MRI reconstruction with models trained at $R=12$ and tested at $R=8$, and vice versa; similarly, we considered CT reconstruction with models trained at $R=8$ and tested at $R=6$, and vice versa. These cases were selected to capture both upward and downward shifts in undersampling factors. Metrics for generalization performance are listed in Table \ref{tab:cross_mri} for MRI, and in Table \ref{tab:cross_ct} for CT. We find that MambaRoll significantly outperforms respective baselines in all cases (p$<$0.05). On average across MRI tasks, MambaRoll offers (PSNR, SSIM) improvements of 3.12dB, 3.88\% over convolutional; 2.19dB, 2.58\% over transformer; and 2.16dB, 2.94\% over SSM baselines. On average across CT tasks, MambaRoll offers (PSNR, SSIM) improvements of 4.15dB, 6.56\% over convolutional; 3.70dB, 4.68\% over transformer; and 2.14dB, 3.65\% over SSM baselines. 

Representative images are depicted in Fig. \ref{fig:ctX}. DD methods exhibit pronounced streaking artifacts, which are generally less severe in PD baselines. MambaRoll delivers the most effective suppression of these residual artifacts. Consistent with quantified performance gains, these results indicate that MambaRoll achieves strong generalization across undersampling factors, outperforming baseline models under mismatched training and testing configurations.

   \begin{figure*}[t]
        \begin{minipage}{0.76\textwidth}
            \centerline{\includegraphics[width=0.98\columnwidth]{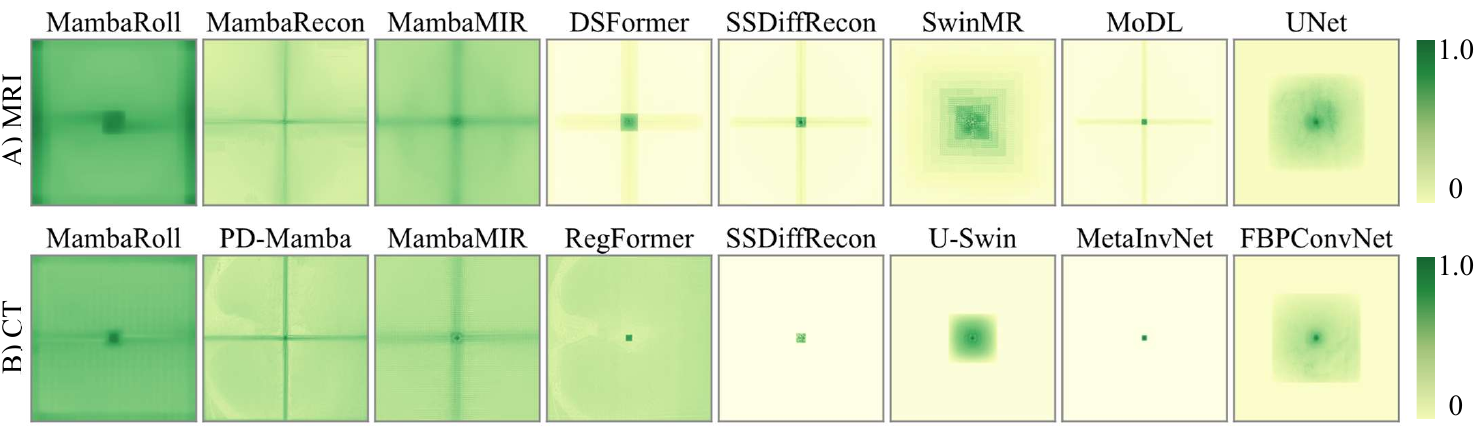}}
        \end{minipage}
        \begin{minipage}{0.23\textwidth}
            \caption{Effective receptive field (ERF) visualizations for MRI (top row) and CT (bottom row) reconstruction. For each method, the corresponding ERF map highlights relative influences of input image regions on the central pixel of the reconstructed image. Green/yellow areas indicate higher/lower contribution, respectively (see colorbar).}
            \label{fig:erfs_mri}
        \end{minipage}
        \hfill
        \vspace{-1mm}
    \end{figure*}

\subsection{Contextual Awareness}
To assess spatial context utilization during reconstruction, we examined contextual awareness—the model's ability to integrate information from spatially distant but contextually dependent regions. In medical image reconstruction, such contextual integration is critical for capturing extended anatomical structures and reducing local ambiguities under undersampled or corrupted measurements. We visualized effective receptive fields (ERFs) by backpropagating gradients from the central pixel of the reconstructed output to the input image, revealing the level at which input regions influence predictions. As depicted in Fig. \ref{fig:erfs_mri}, MambaRoll exhibits the broadest and most coherent ERF coverage, indicating consistent multi-scale contextual integration. Conventional SSMs show moderately expansive but focalized ERFs along horizontal and vertical lines, reflecting raster-scan biases used to sequentialize image pixels. Transformer-based models display relatively compact or fragmented ERFs —especially those using local window attention (e.g., SwinMR, USwin)— suggesting that despite their theoretical capacity for global attention, practical implementations may underutilize broad spatial cues. CNN-based models show variable ERFs: deeper models with UNet-like structures (e.g., UNet, MetaInvNet) achieve moderately spread fields through multi-resolution processing and skip connections, while uniform designs maintaining consistent feature map resolution (e.g., MoDL, FBPConvNet) remain highly localized. These findings highlight that by combining autoregressive modeling with multi-scale state space processing, MambaRoll more coherently captures long-range dependencies, providing broad spatial sensitivity valuable for anatomical coherence and global structure recovery in medical image reconstruction.

\begin{table}[t]
  \centering
  \footnotesize
    \caption{Average run time (msec) and memory load (MB) per cross-section for MRI reconstruction during training and inference phases.}
  \renewcommand{\arraystretch}{1.1}
    \resizebox{0.725\columnwidth}{!}{
      \begin{tabular}{lcccc}
          \hline
            & \multicolumn{2}{c}{Run time (msec)} & \multicolumn{2}{c}{Memory (MB)} \\
            \cmidrule(lr){2-3} \cmidrule(lr){4-5} 
            & Training & Inference & Training & Inference \\ \hline
            UNet & 42.37 & 11.56 & 1516 & 550 \\ \hline
            MoDL & 50.60 & 14.99 & 1398 & 304 \\ \hline
            SwinMR & 149.60 & 49.85 & 3864 & 870 \\ \hline
            SSDiffRecon & 196.73 & 339.40 & 3858 & 740 \\ \hline
            DSFormer & 750.21 & 248.89 & 16090 & 1338 \\ \hline
            MambaMIR & 121.34 & 36.84 & 2382 & 474 \\ \hline
            MambaRecon & 139.89 & 36.55 & 2304 & 462 \\ \hline
            MambaRoll & 96.09 & 29.67 & 2568 & 358 \\ \hline
          \end{tabular}
   }
  \label{tab:compute_mri}
\end{table}

\begin{table}[t]
  \centering
  \footnotesize
    \caption{Average run time (msec) and memory load (MB) per cross-section for CT reconstruction during training and inference phases.}
  \renewcommand{\arraystretch}{1.1}
   \resizebox{0.725\columnwidth}{!}{
      \begin{tabular}{lcccc}
          \hline
            & \multicolumn{2}{c}{Run time (msec)} & \multicolumn{2}{c}{Memory (MB)} \\
            \cmidrule(lr){2-3} \cmidrule(lr){4-5} 
            & Training & Inference & Training & Inference \\ \hline
            FBPConvNet & 84.93 & 13.23 & 1962 & 974 \\ \hline
            MetaInvNet & 77.57 & 29.16 & 5144 & 3288 \\ \hline
            USwin & 246.06 & 80.59 & 5926 & 1082 \\ \hline
            SSDiffRecon & 193.19 & 346.74 & 7076 & 3866 \\ \hline
            RegFormer & 108.64 & 48.20 & 7890 & 5884 \\ \hline
            MambaMIR & 143.05 & 41.29 & 2830 & 532 \\ \hline
            PD-Mamba & 303.37 & 90.93 & 9264 & 4244 \\ \hline
            MambaRoll & 134.23 & 49.79 & 6144 & 3374 \\ \hline
          \end{tabular}
   }
  \label{tab:compute_ct}
\end{table}

    \begin{table}[t]
        \centering
        \footnotesize
            \caption{Performance of MambaRoll variants, where `w/o PD-SSM' ablates PD-SSM modules, `w/o AR' ablates autoregressive processing, `w/o comp. SSM' ablates compressed SSM blocks, and `w/o DC' ablates DC modules.} 
         \renewcommand{\arraystretch}{1.1}
          \resizebox{0.9\columnwidth}{!}{%
        \begin{tabular}{lccccc}
        \hline
        \multirow{2}{*}{} & \multicolumn{2}{c}{fastMRI - FLAIR, $R = 4$} & \multicolumn{2}{c}{CT, $R = 4$} \\ \cline{2-5} & PSNR & SSIM & PSNR & SSIM \\ \hline
        MambaRoll 
            & \textbf{43.86$\pm$3.32} & \textbf{98.43$\pm$1.43}  
            & \textbf{40.72$\pm$3.56} & \textbf{94.05$\pm$4.86} \\ \hline
        
        w/o PD-SSM
            & 40.49$\pm$2.70 & 97.08$\pm$1.98 
            & 31.00$\pm$1.27 & 83.50$\pm$5.00 \\ \hline

        w/o AR
            & 42.62$\pm$3.11 & 98.01$\pm$1.66
            & 37.11$\pm$2.56 & 90.53$\pm$5.33 \\ \hline  

        w/o comp. SSM
            & 41.43$\pm$3.11 & 97.44$\pm$2.13 
            & 35.65$\pm$2.41 & 88.51$\pm$6.17 \\ \hline

        w/o DC
            & 36.88$\pm$2.50 & 93.44$\pm$3.38
            & 38.66$\pm$2.99 & 92.14$\pm$5.15 \\ \hline

        \end{tabular}
        }
        \label{tab:ablation}
    \end{table}

\subsection{Computational Complexity}
Computational complexity is an important consideration for adoption of learning-based models. Run times and memory loads for competing methods are listed in Table \ref{tab:compute_mri} for MRI and Table \ref{tab:compute_ct} for CT. CNN-based methods generally demonstrate fast run times and low memory demand, though with exceptions. In CT reconstruction, MetaInvNet shows higher compute load as a PD architecture due to intrinsic complexity of Radon transformation within data-consistency modules. Transformer-based methods typically incur elevated compute burden with notably higher run times and memory demand despite efficiency-oriented implementations (e.g., windowed attention mechanisms). While RegFormer has competitive run times to CNNs in CT reconstruction, it has high memory demand as a PD architecture. SSM-based methods generally demonstrate efficiency levels between CNNs and transformers, with MambaRoll showing the highest computational efficiency among SSMs, achieving run times and memory demand closest to CNN methods.
Comparative visualizations of performance and efficiency are presented in Fig. \ref{fig:compute}. While some CNN-based methods achieve higher computational efficiency (lower latency and/or memory load), this comes at the expense of reconstruction quality (lower PSNR and SSIM). MambaRoll offers an attractive balance between reconstruction performance and computational efficiency. For CT reconstructions, PD models (PD-Mamba, RegFormer, SSDiffRecon, MetaInvNet) incur additional computational costs due to forward and inverse Radon transformations in data-consistency modules, effectively addressed by MambaRoll's efficient design. These results suggest particular promise for MambaRoll in achieving high fidelity and efficiency in medical image reconstruction.

\subsection{Ablation Studies}

\subsubsection{Network Architecture}
We conducted ablation studies to evaluate major architectural elements of MambaRoll. To assess PD-SSM modules, we trained a `w/o PD-SSM' variant that removed the PD-SSM modules albeit retained the refinement modules so as to create an unrolled convolutional architecture. To assess autoregressive modeling across spatial scales, we trained a `w/o AR' variant that used PD-SSM modules operating at a single spatial scale retaining input resolution. To assess SSM layers in capturing contextual representations, we trained a `w/o comp. SSM' variant that ablated compressed SSM blocks from PD-SSM modules. To assess DC blocks, we trained a `w/o DC' variant that ablated DC blocks albeit retaining their residual connections to create a DD SSM model. Table \ref{tab:ablation} lists performance metrics for variants. MambaRoll yields superior performance against all variants, corroborating the importance of key architectural design elements.

\subsubsection{Loss Function}
\begin{figure}[t]
    \centering
    \includegraphics[width=0.7\linewidth]{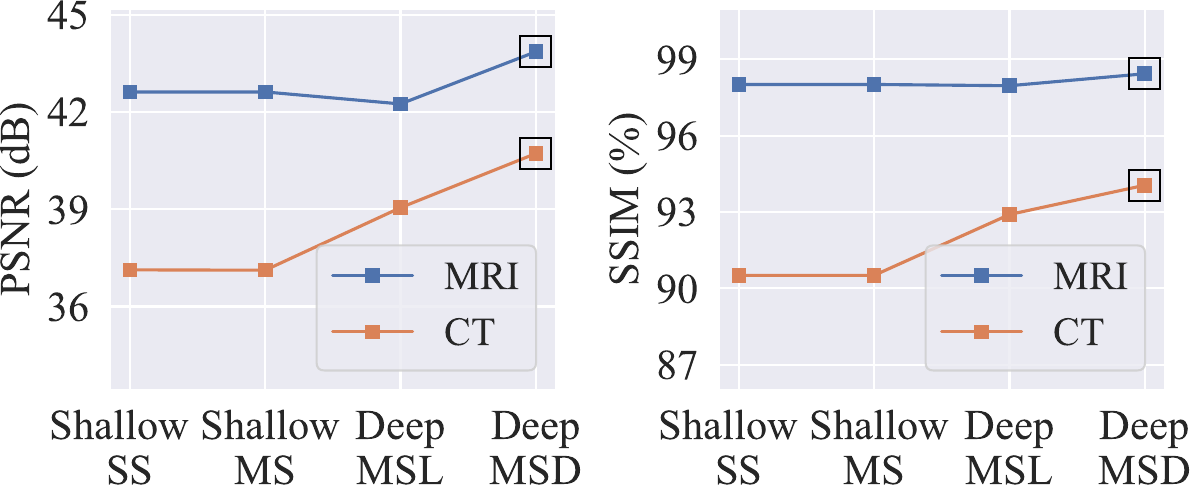}
    \caption{Performance (FLAIR, $R=4$ for MRI and $R=4$ for CT) with the proposed DMSD loss (marked with squares), along with Shallow SS, Shallow MS, and Deep MSL variants.}
    \label{fig:ablation_loss}
\end{figure}

We evaluated the importance of DMSD loss on reconstruction performance by training MambaRoll variants with alternative supervision. A `Shallow SS' variant only used a single-scale loss term on the network output. A `Shallow MS' variant used the single-scale loss term on the network output alongside multi-scale loss terms on spatially downsampled versions of the output (with a total of $S$ scales). A `Deep MSL' variant used a deep multi-scale loss on the latent feature maps extracted by PD-SSM modules (i.e., $g_s$), instead of the decoded images as in MambaRoll. Figure \ref{fig:ablation_loss} shows that the proposed DMSD loss yields superior performance against variants, corroborating the importance of the proposed training objective for autoregressive learning.

\begin{figure}[t]
    \centering
    \includegraphics[width=0.7\linewidth]{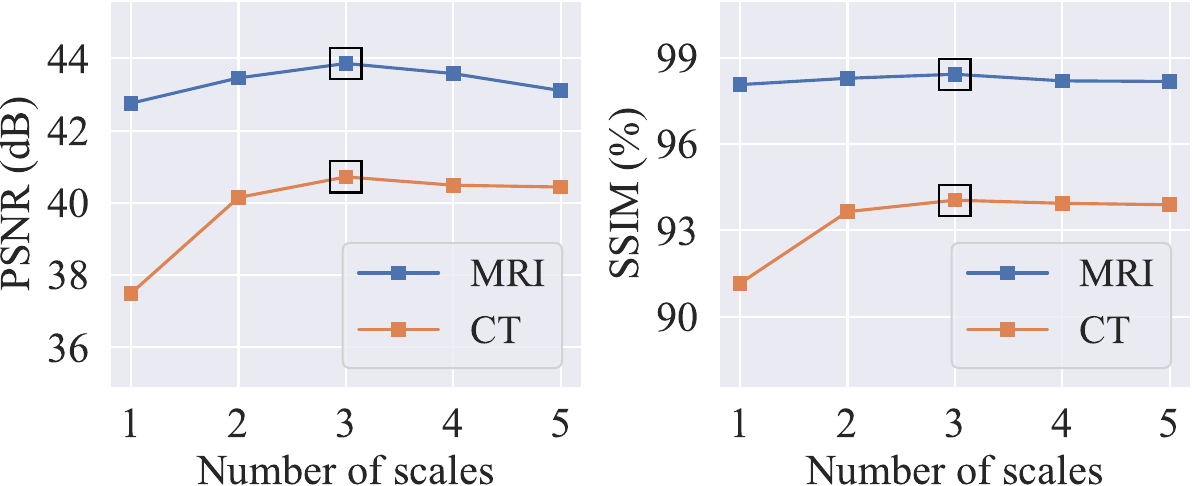}
    \caption{Performance of MambaRoll variants that prescribed varying numbers of spatial scales (selected values marked with squares).}
    \label{fig:ablation_scale}
\end{figure}

\begin{figure}[t]
    \centering
    \includegraphics[width=0.8\linewidth]{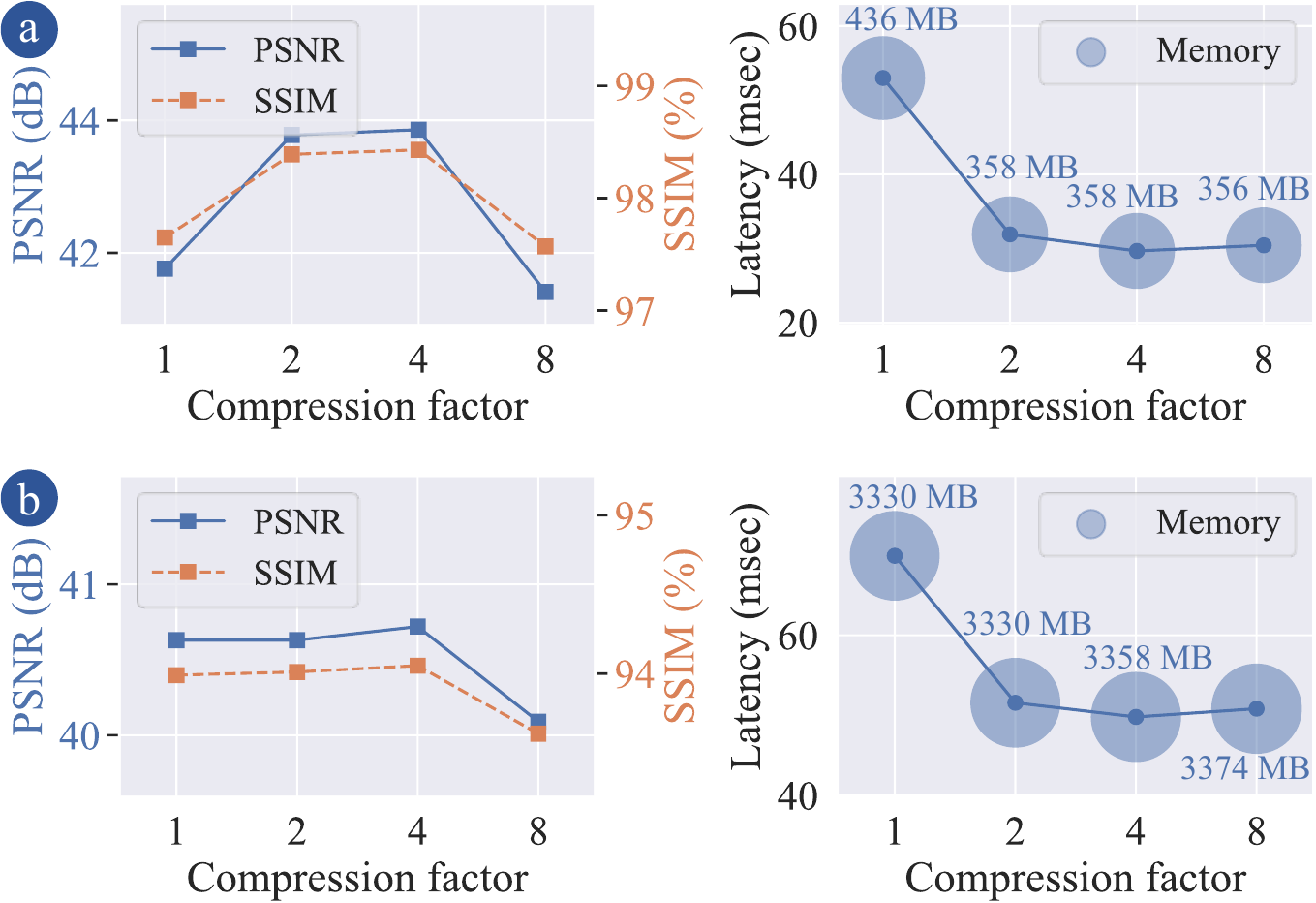}
    \caption{Performance of MambaRoll variants that prescribed varying compression factors \textbf{(a)} on FLAIR, $R=4$ for MRI; \textbf{(b)} on $R=4$ for CT. PSNR, SSIM are displayed in the left panel; memory load, latency are displayed in the right panel.}
    \label{fig:ablation_shuffle_mri}
\end{figure}

\subsubsection{Design Parameters}
We assessed the influence of spatial scales for autoregressive modeling and compression factors in PD-SSM modules. As shown in Fig. \ref{fig:ablation_scale}, performance increases with growing scales until $S=3$, without apparent benefits beyond this point, indicating a near-optimal performance-efficiency trade-off. Meanwhile, Fig. \ref{fig:ablation_shuffle_mri} shows performance and efficiency across different compression factors in compressed SSM blocks. While larger compression factors enhance contextual information exchange across distant regions by increasing spatial sampling steps, they expand feature maps along channel dimensions, elevating computational overhead. A factor of $J=4$ provides optimal performance while maintaining efficiency. Factors beyond 4 degrade performance without efficiency benefits, likely due to excessive information dispersion disrupting local spatial coherence necessary for effective medical image reconstruction.

\section{Discussion}
The utility of MambaRoll could be further improved by addressing several technical limitations. A first group of improvements concern the learning setup. Here, a supervised setup using fully-sampled acquisitions as ground truth was considered. To enable training on datasets with only undersampled data, self-supervised learning could be adopted \cite{yaman2020}. When curating any training data is difficult altogether, test-time training could be used for scan-specific reconstruction at increased inference cost \cite{darestani2021accelerated,korkmaz2022unsupervised}.

A second group of improvements concern learning objectives. Here we observed high-quality reconstructions using models trained for autoregressive image prediction guided via DMSD loss. Our results suggest that autoregressive modeling might be a promising alternative to generative methods based on adversarial and diffusive losses \cite{tian2024VAR}. Yet, combining autoregressive loss with other generative losses might help further improve sensitivity to fine details. While our primary focus here was on architectural design of reconstruction models, MambaRoll can be combined with generative learning objectives to examine their potential benefits in future work. 

A third group of improvements concern reconstruction tasks. Here we built models that operate on single-modality acquisitions (a single MRI contrast or a single CT contrast). Recent work suggests SSMs can also aid in joint reconstruction of multiple modalities \cite{mmrmamba}. A basic approach to extend MambaRoll for multi-modal reconstruction would be to assign separate channels per modality in the input-output layers and across DC layers in PD-SSM modules \cite{xiang2019rec}. Separate model branches could also be employed to process each modality, while cross-modality interactions are mediated by attentional or state-space modules for fine-grained control of feature representations shared between modalities \cite{mmrmamba}.

A final group of improvements concerns architectural design. Each PD-SSM module integrates a linear SSM layer between scale-specific convolutional layers that downsample/upsample feature maps, enabling autoregressive modeling across spatial scales. Our experiments suggest that this design outperforms existing architectures based on conventional Mamba modules (e.g., MambaMIR, MambaRecon), which use a popular gating mechanism to modulate SSM outputs \cite{liu2024vmamba}. A hybrid architecture that pools linear and gated SSMs might offer performance benefits. Dual-domain approaches that combine spatial and spectral processing can also enhance performance \cite{KikiNet}. Further work is needed to systematically explore the benefits of such hybrid architectures.

\section{Conclusion}
We introduced a novel deep learning method for medical image reconstruction, MambaRoll, based on physics-driven autoregressive state-space modeling. MambaRoll leverages autoregressive prediction across multiple spatial scales along with physics-driven SSM modules to effectively aggregate contextual features. To enhance learning efficacy, MambaRoll employs a deep multi-scale decoding loss tailored to the autoregressive prediction task. Demonstrations on MRI and CT reconstructions indicate that the proposed model consistently outperforms state-of-the-art methods in terms of image quality, without compromising computational efficiency. Therefore, MambaRoll holds great potential for improving the utility of learning-based medical image reconstruction.

\bibliographystyle{IEEETran} 
\bibliography{IEEEabrv,refs}

\end{document}